\documentclass{PoS}
\let\ifpdf\relax
\usepackage{wrapfig}
\newcommand\beq{\begin{eqnarray}}
\newcommand\eeq{\end{eqnarray}}
\newcommand {\bmp}{\begin{minipage}}
\newcommand {\emp}{\end{minipage}}
\newcommand \reffig {Fig.~\ref}
\newcommand \refeq {Eq.~\ref}

\newcommand\bfn{{\bf n}}
\newcommand\bfmu{{\bf \mu}}
\newcommand\bfnu{{\bf \nu}}

\title{Large $N_c$ volume reduction and chiral random matrix theory}

\ShortTitle{Large $N_c$ volume reduction and chiral random matrix theory}

\author{\speaker{Jong-Wan Lee}\\
	KEK Theory Center, Institute of Particle and Nuclear Studies, 
	High Energy Accelerator Research Organization (KEK), Tsukuba 305-0801, Japan\\
        E-mail: \email{jongwan@post.kek.jp}}

\author{Masanori Hanada\\
	KEK Theory Center, Institute of Particle and Nuclear Studies, 
	High Energy Accelerator Research Organization (KEK), Tsukuba 305-0801, Japan\\
        E-mail: \email{hanada@post.kek.jp}}

\author{Norikazu Yamada\\
	KEK Theory Center, Institute of Particle and Nuclear Studies, 
	High Energy Accelerator Research Organization (KEK), Tsukuba 305-0801, Japan\\
        E-mail: \email{norikazu.yamada@kek.jp}}

\abstract{
Motivated by recent progress on the understanding of the Eguchi-Kawai (EK) volume equivalence and growing interest in conformal window, we simultaneously use the large-$N_c$ volume reduction and Chiral Random Matrix Theory ($\chi$RMT) to study the chiral symmetry breaking of four dimensional SU($N_c$) gauge theory with adjoint fermions in the large $N_c$ limit. 
Although some cares are required because the $\chi$RMT limit and 't Hooft limit are not compatible in general, we show that the breakdown of the chiral symmetry can be detected in large-$N_c$ gauge theories. 
As a first step, we mainly focus on the quenched approximation to establish the methodology. 
We first confirm that heavy adjoint fermions, introduced as {\it the center symmetry preserver}, work as expected and thanks to them the volume reduction holds. 
Using massless overlap fermion as a probe, we then calculate the low-lying Dirac spectrum for fermion in the adjoint representation to compare to that of $\chi$RMT, and find that chiral symmetry is indeed broken in the quenched theory.
}
\FullConference{The XXXth International Symposium on Lattice Field Theory - Lattice2012\\
		July 24-29, 2012\\
		Cairns, Australia}

\begin{document}

\section{Introduction}
\label{sec:introduction}

In modern theoretical particle physics, numerical simulations of quantum field theory on the lattice play an important role in non-perturbative studies of strongly coupled gauge theories, e.g. quantum chromodynamics (QCD), from the first principle. 
However, lattice calculations often encounter practical challenges such as large finite-volume effects and sign problems. 
For instance, lattice studies of the infrared behavior of conformal theories are extremely difficult due to finite size of the lattice. 
These challenges may be overcome by constructing lattice formulations for large-$N_c$ gauge theories (for a review, see \cite{Panero}) or for effective field theories \cite{Chen}, where various nice properties hold.
In particular, the large-$N_c$ volume equivalence \cite{EK} has recently received large attention because of the theoretical understandings of the center symmetry stabilization and the success of the numerical tests \cite{Narayanan2003,Kovtun2007,Bringoltz2009,Azeyanagi,Arroyo}.\footnote{Unbroken center symmetry is the most crucial condition for the large-$N_c$ volume equivalence. The original proposal for pure $U(N_c)$ gauge theory by Eguchi-Kawai failed due to spontaneous breaking of center symmetry \cite{Bhanot1982}.}

One of the motivation of our work stems from the growing interest in conformal window and the walking technicolor model (WTM), where conventional lattice QCD techniques can be easily extended to QCD-like theories. 
A candidate of the minimal WTM is two-flavor $SU(2)$ gauge theory with fermions in the adjoint representation \cite{Sannino}. 
Bearing this in mind, we consider numerical calculations of $SU(N_c)$ gauge theories with $n_f=2$ adjoint fermions in the large $N_c$ limit.
We first clarify the difference between the $\chi$RMT limit and 't Hooft large-$N_c$ limit in large-$N_c$ gauge theories. Then, we establish the methodology - the way to demonstrate the spontaneous chiral symmetry breaking (S$\chi$SB) by applying the $\chi$RMT techniques and the large-$N_c$ volume equivalence to large-$N_c$ gauge theories (for earlier work along this direction, see \cite{Narayanan2004,Hietanen}). As a concrete example, we study the heavy adjoint two flavor QCD up to $N_c=16$ on a $2^4$ lattice, which approximates the quenched QCD, and find that the chiral symmetry is indeed broken.  

\section{The 't Hooft versus $\chi$RMT limit in large-$N_c$ gauge theory}
\label{largeNc}

In large-$N_c$ gauge theories, the 't Hooft limit is the large-$N_c$ limit in which the 't Hooft coupling $\lambda = g^2 N_c$ and the number of quark flavor $n_f$ are fixed \cite{Hooft}. The space-time volume $V$ and the quark mass $m$ are also fixed. In this limit, the theory is dramatically simplified, i.e. the $1/N_c$ expansion has a natural topological structure and a certain class of Feynmann diagrams, so called "$planar$ $diagrams$", provide dominant contributions in the perturbation theory. For example, the vacuum expectation value of a properly normalized single trace operator $\hat{\mathcal{O}}$ can be expanded as,
\beq
\langle \hat{\mathcal{O}} \rangle = \sum_{h,B=0}^\infty c_{h,B} (\lambda, V, m) N_c^{-2h-B},
\label{single_trace_Nc}
\eeq
where $h$ and $B$ are the number of handles and boundaries of diagrams. The connected correlation functions of more than one operators have the same structure. 

In the 't Hooft limit, various nice properties hold; in particular, vacuum expectation values of products of gauge-invariant operators are factorized up to $O(1/N_c)$ corrections, $\langle\mathcal{O}_1\mathcal{O}_2\rangle=\langle\mathcal{O}_1\rangle \langle\mathcal{O}_2\rangle+O(1/N_c)$. An important consequence of the large-$N_c$ factorization is the so-called Eguchi-Kawai (EK) volume equivalence \cite{EK}; provided center symmetry is not broken spontaneously, the Wilson loop amplitudes in an arbitrarily small volume agree with those of the usual large-volume lattice gauge theory.  
The large-$N_c$ volume equivalence can also be understood as an example of orbifold equivalence \cite{Kovtun2007} and the general statement is as follows. Starting with an original (parent) theory, we obtain a new (daughter) theory by performing a projection under some discrete subgroup of the global symmetry of the parent theory. If the discrete symmetry does not break down spontaneously, the correlation functions of invariant (neutral) sectors of operators in both theories are equal up to a trivial rescaling factor. 

Provided the chiral symmetry is spontaneously broken, the dynamics of QCD at below $\Lambda_{QCD}$ can be described by the low-energy effective theory where the degrees of freedom are Goldston bosons such as pions instead of quarks and gluons. If we go further inside the $\epsilon$-regime, where the one-fourth of space-time volume is much smaller than the pion Compton length, the relevant degrees of freedom are zero modes of pions and the density of Dirac eigenvalues near zero is related to the chiral condensate $\Sigma$ via Banks-Casher relation \cite{Banks}. The distribution of the low-lying Dirac eigenvalues can be calculated using the chiral random matrix theory ($\chi$RMT) in the microscopic limit, $N\rightarrow \infty$ while having $mN$ fixed, where the matrix size $N$ is identified by $V$ in the standard gauge theory (for a review, see \cite{Verbaarschot}). 
In large-$N_c$ gauge theories, the number of color charges $N_c$ are also relevant degrees of freedom and the corresponding limit may be obtained by taking the large-$N_c$ limit with a fixed $m N \sim m V N_c^\alpha$, where the constant $\alpha$ can depend on the representation of fermions. Let us call this limit as the "$\chi$RMT limit" which is different from the 't Hooft limit: the fermion mass scales with $N_c$ and hence $c_{h,B}$ in \refeq{single_trace_Nc} also has a nontrivial $N_c$-dependence implying that the 't Hooft power counting rules and the large-$N_c$ equivalences are not valid. 
In the next section, however, we argue that the EK equivalence and the $\chi$RMT combined with lattice simulations can be used to detect the S$\chi$SB in QCD-like theories with some cares.

\section{Strategy of detecting the spontaneous chiral symmetry breaking}
\label{sec:strategy}

In this section we establish the way to use the $\chi$RMT in large-$N_c$ gauge theory for detection of the S$\chi$SB.
First let us recall how one can confirm the chiral symmetry breakdown in the ordinary $SU(3)$ QCD.
The criterion for the S$\chi$SB is the nonzero chiral condensate in the standard thermodynamic limit,
\begin{center}
$\langle\bar{\psi}\psi\rangle\neq 0$ in the massless limit $m\to 0$ after taking the large-volume limit $V\to\infty$.
\end{center}
On the other hand, for QCD in the $\epsilon$-regime, the S$\chi$SB is recognized by the agreement of the low-lying Dirac spectrum with the prediction from $\chi$RMT.

In the case of large-$N_c$ gauge theory, the logic is as follows. First, we calculate the low-lying Dirac spectrum of the large-$N_c$ gauge theory in a small box (e.g. on a $1^4$ or $2^4$ lattice) and compare the spectrum with the prediction from $\chi$RMT. 
Here the probe mass $m_{probe}$ must scale as $m_{probe}\sim 1/N_c^\alpha$ for large $N_c$, otherwise such that the $\chi$RMT limit is realized. Because of this scaling, the EK equivalence, requiring $N_c \rightarrow \infty$ with $m$ fixed, can not be applied directly. 
However, following the same logic as the QCD case except that $V$ is now replaced with $N_c^\alpha$, the agreement of the spectrum with the prediction from $\chi$RMT still establish the S$\chi$SB of the large-$N_c$ gauge theory in a small box. 
Then the EK equivalence leads to the S$\chi$SB of the large-$N_c$ gauge theory at large volume. 
The power $\alpha$ is determined so that $m_{probe}$ and the Dirac eigenvalues near zero have the same $N_c$ -dependence, and we will see $\alpha=1$ in our setup. 
The schematic diagram of the detection of the S$\chi$SB in the large $N_c$ limit is shown in Fig. 1.

\begin{figure}
\begin{center}
\includegraphics[width=.8\textwidth]{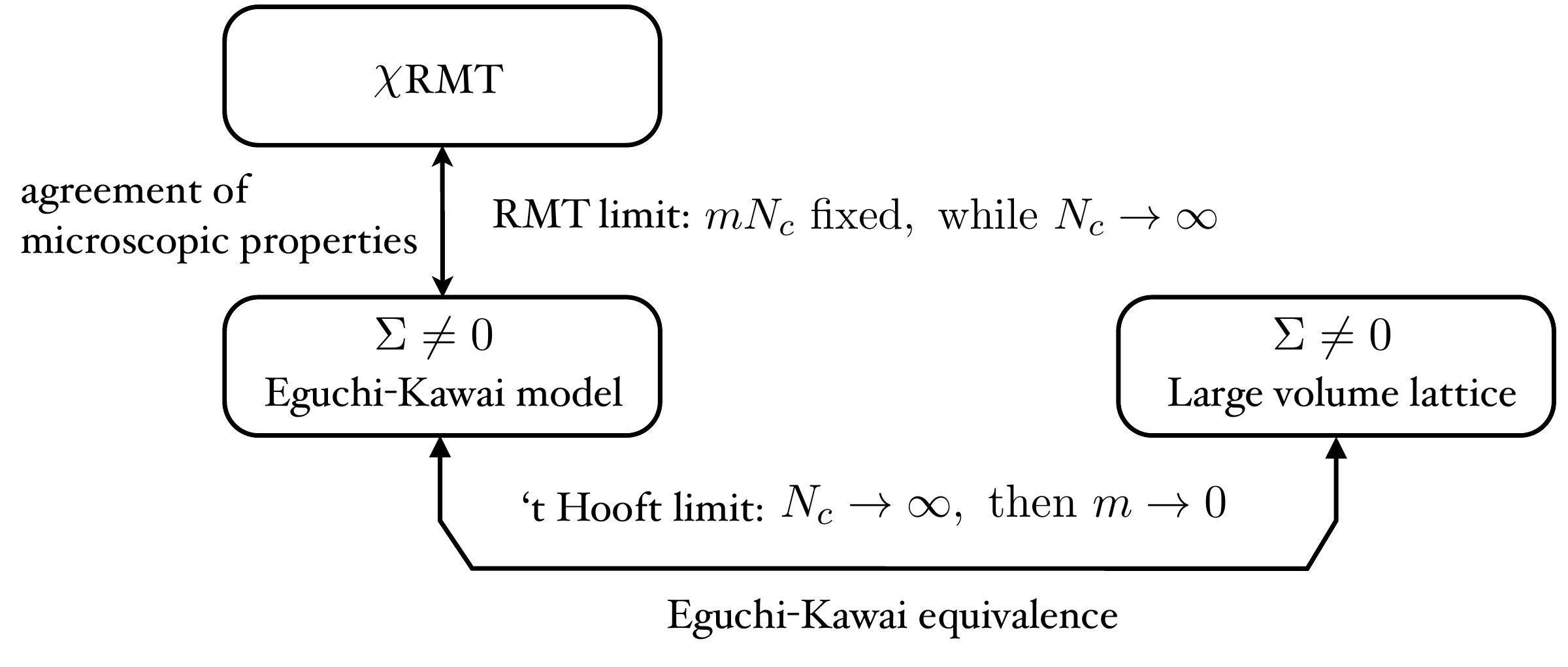}
\caption{%
\label{detection_chi_sym}
Detection of the spontaneously breakdown of chiral symmetry in large-$N_c$ gauge theories using the $\chi$RMT and EK equivalence.
}
\end{center}
\end{figure}

\section{Numerical simulation of the Eguchi-Kawai model}
Numerical simulations are performed with the Wilson gauge action and Wilson-Dirac fermions in the adjoint representation on a $2^4$ lattice.
\beq
S_g = 2 N_c^2 b \sum_\bfn \sum_{\bfmu<\bfnu} \left(1-\frac{1}{N_c} \textrm{Re} \textrm{Tr} P_{\mu\nu} (\bfn)\right), ~~~\bfn \in 2^4
\eeq
and
\beq
S_f = \sum_{j=1}^2 \sum_\bfn \overline{\psi}_{\bfn,j}\left(\psi_{\bfn,j}-\kappa \sum_\mu^4 \left[(1-\gamma_\mu)U_{\bfn,\mu}^{adj} \psi_{\bfn+\mu,j}
+(1+\gamma_\mu)U_{\bfn-\mu,\mu}^{\dagger,adj} \psi_{\bfn-\mu,j}\right]\right),
\eeq
where $b$ and $\kappa$ represent the inverse of the 't Hooft coupling, $b=1/g^2 N_c$, and the hopping paremeter which is related to bare quark mass by $\kappa=(2ma+8)^{-1}$, respectively. The Plaquettes $P_{\mu\nu}$ are built of link variables in the fundamental representation as usual. For the fermionic action, the link variables in the adjoint representation are defined by
\beq
U_{a,b}^{adj}=\frac{1}{2}\textrm{Tr}[T^a_F U T^b_F U^\dagger],
\eeq
where $T^a_F$ are $SU(N_c)$ generators in the fundamental representation.
This action is invariant under $SU(N_c)$ local gauge transformation as well as $\mathbb{Z}_{N_c}$ global center transformation:
\beq
U_{\bfn,\bfmu}~ \longrightarrow ~\Omega_{\bfn} U_{\bfn,\bfmu} \Omega_{\bfn+\bfmu}^\dagger, ~~\Omega_{\bfn} \in SU(N_C),
~~
\textrm{and}~~
U_{\bfn,\bfmu}~ \longrightarrow ~ e^{2\pi i \bfn_{\bfmu} /N_c} U_{\bfn,\bfmu}, ~~\bfn_{\bfmu} \in \mathbb{Z}_{N_c}.
\eeq
Throughout our studies, we use periodic boundary conditions for all lattice directions in both link variables and fermion fields.

Our lattice simulations consist of two parts: 1) quenched calculations ($\kappa=0$ or equivalently $ma$ is infinite), as a nontrivial check of our numerical code by confirming the breaking of center symmetry at weak coupling, 2) simulations for two adjoint fermions whose mass is of order $1/a$ ($\kappa=0.09$) \footnote{These heavy adjoint fermions serves as a "{\it center symmetry preserver}", where our chose of $\kappa=0.09$ and $b=0.5$ is inside of the conjectured center symmetric region in the $\kappa-b$ plane \cite{Bringoltz2012}, but do not play any role in other parts of dynamics. Thus the case 2) can be also considered as a quenched calculation.}, where low-lying Dirac eigenvalues are calculated by using a massless overlap-Dirac fermion as a probe. 
We performed simulations at $b=0.5$ for up to $N_c=16$ which is relatively smaller than that used for simulations of a single-site EK model \cite{Bringoltz2009,Azeyanagi,Bringoltz2012}. As we will see below, however, we could obtain good large $N_c$ limits since we have additional suppression of the finite volume effects thanks to the larger volume $V=2^4$. For $N_c=8$ and $\kappa=0$, we also performed simulations at $b=0.3$ and $0.4$ corresponding to the strong and intermediate couplings, respectively.
For all lattice simulations, we used the Hybrid Monte Carlo (HMC) algorithm; $200$ trajectories are used for the initial thermalization, typically $500$ configurations are generated for each ensemble, and every two adjacent configurations are separated by $10$ trajectories.

\subsection{center symmetry}

In order to use the large-$N_c$ volume equivalence, it is essential to confirm that the center symmetry is unbroken. 
Some evidences of which center symmetry is unbroken are as follows: (1) the Polyakov line $P_\mu$ scatters radially in the vicinity of origin in the complex plane, (2) the magnitude of $P_\mu$ approaches zero as $N_c$ increases where the predicted asypmtotic scaling behavior is $1/N_c$, (3) the average plaquette value measured from the reduced model agrees with that measured from the large-volume lattice gauge theory. 
The Polyakov line along $\mu$-direction in a $2^4$ lattice is defined by
\begin{eqnarray}
P_{\mu}({\bf x})
=
\frac{1}{N_c} 
\textrm{Tr} U_{\mu}({\bf x})U_{\mu}({\bf x}+e_\mu). 
\end{eqnarray}
\begin{figure}
\begin{center}
\includegraphics[width=.32\textwidth]{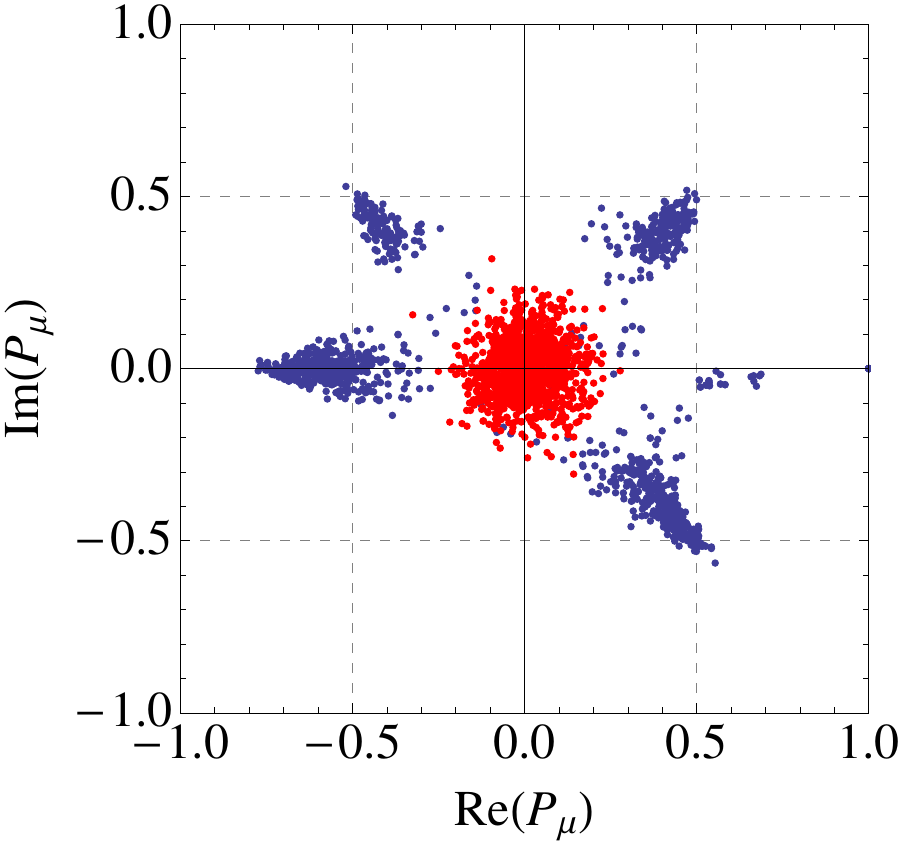}
\hskip .7in
\includegraphics[width=.45\textwidth]{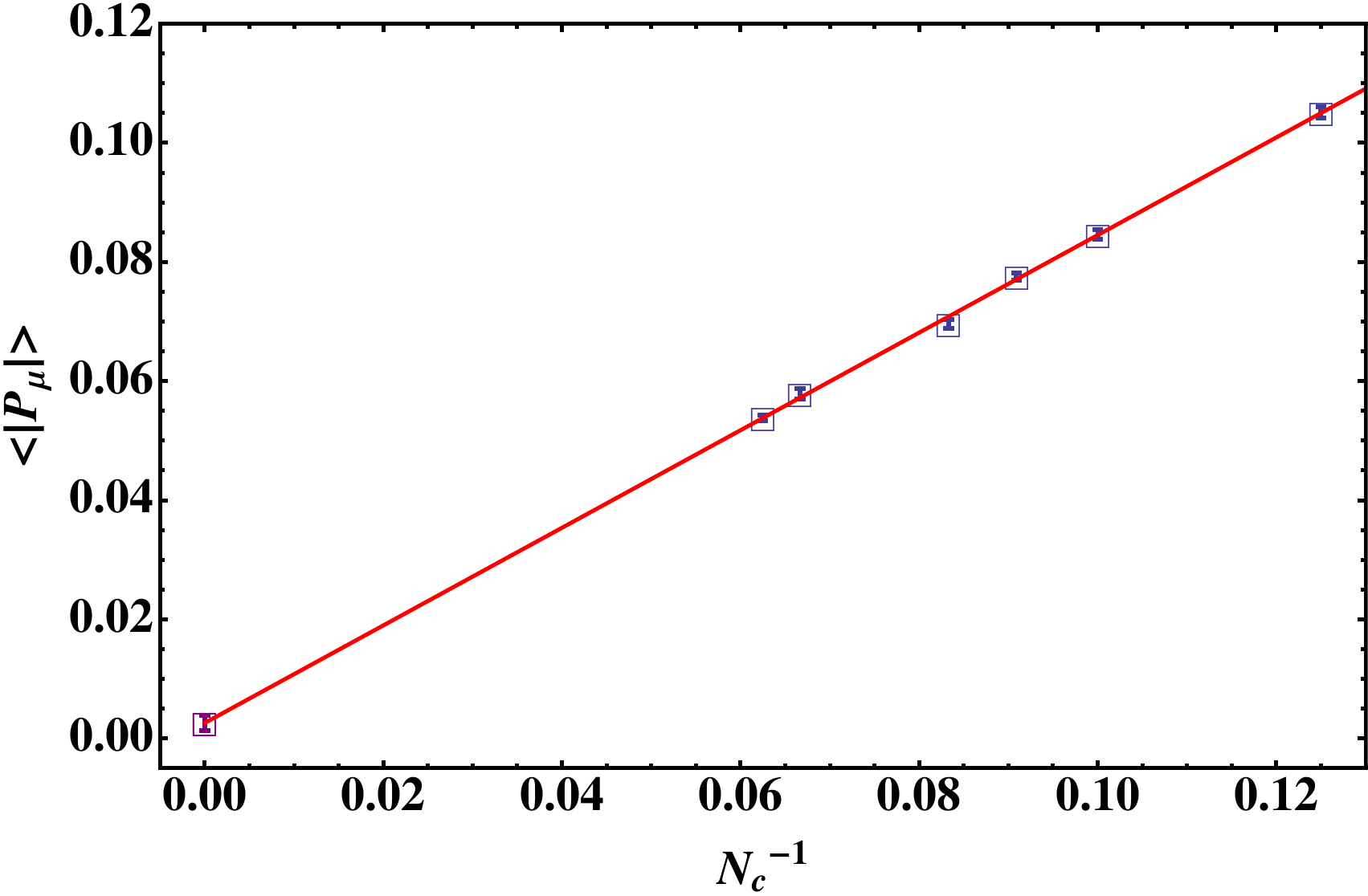}
\caption{%
\label{centersymmetry}%
(Left) Scatter plots of the Polyakov lines, which are overaged over all $2^3$ sites orthogonal to the line dirction $\mu=1$, for $N_c=8$ and $b=0.5$.  The red and blue dots are about $150$ data points obtained from ensembles for $\kappa=0.09$ and $\kappa=0$, respectively. 
(Right) Ensemble averaged values of magnitudes of the Polyakov lines versus $N_c^{-1}$ for the EK model coupled to two heavy adjoint fermioins. The red solid line represent the fits of the $5\leq N_c \leq 16$ data to the function  $c_0+c_1/N_c$.}
\end{center}
\end{figure}
From numerical simulations for quenched EK model with various values of $N_c$ and $b$, we first reproduced the well known results for pure Yang-Mills theory: the center symmetry is unbroken at strong coupling, but broken at weak coupling. 
In particular, we show scatter plots of the $P_\mu$'s for $N_c=8$ and $b=0.5$ in \reffig{centersymmetry}, where the red and blue dots represent the data points for $\kappa=0.09$ (EK model coupled to two heavy adjoint fermions) and $\kappa=0$ (quenched EK model), respectively. 
The two colors have very different behaviours: the red develops a cluster around the origin, which is consistent with the senario of unbroken center symmetry, while the blue is localized at some of the elements of the center of $SU(N_c)$, which implies that the center symmetry is broken. 
At strong coupling, i.e. $b=0.3$, the $P_\mu$'s radially scatter around the origin in both simulations. 
We also calculated ensemble averages of $|P_\mu|$ for the case of $\kappa=0.09$ and $b=0.5$. As shwon in \reffig{centersymmetry}, they are linear in $1/N_c$ and extrapolate to zero at $N_c=\infty$. Finally, we caculated the average plaquette values and performed extrapolations to $N_c=\infty$. The values we obtained are $0.72053(71)$ for $\kappa=0.09$ and $0.72733(12)$ for $\kappa=0$. Compared with the large-volume lattice simulation \cite{Bringoltz2009} and the single-site EK model for quenched QCD \cite{Azeyanagi}, the former is in good agreement while the latter is systematically larger. In summary, the center symmetry is intact in the theory with two heavy adjoint fermions and thus the EK volume equivalence holds; the euqivalent large-volume lattice theory approximates the quenched large-$N_c$ QCD.

\subsection{Dirac spectrum and comparison to $\chi$RMT}
\label{sec:dirac_spectrum}

\begin{figure}
\begin{center}
\includegraphics[width=.45\textwidth]{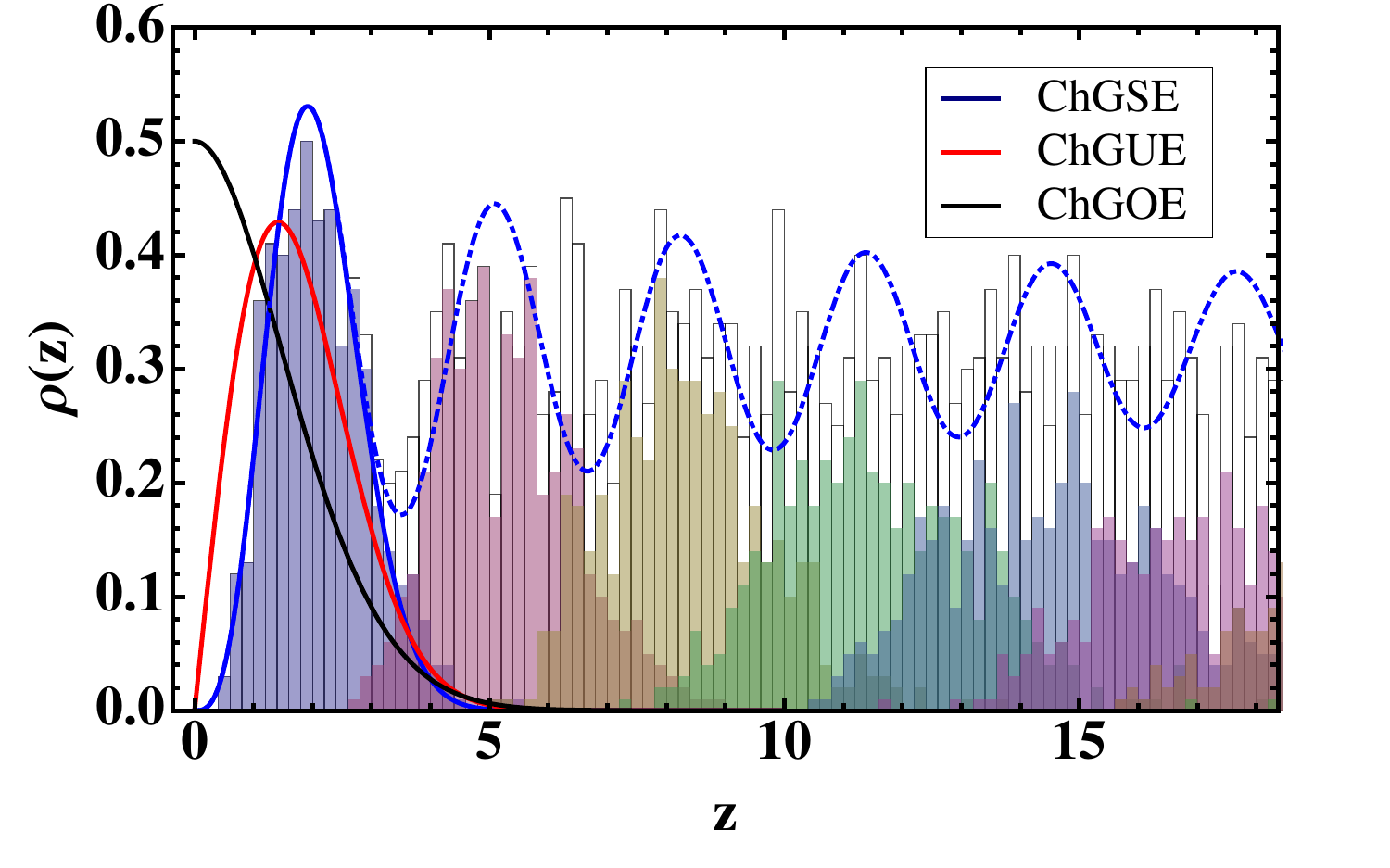}
\hskip .2in
\includegraphics[width=.45\textwidth]{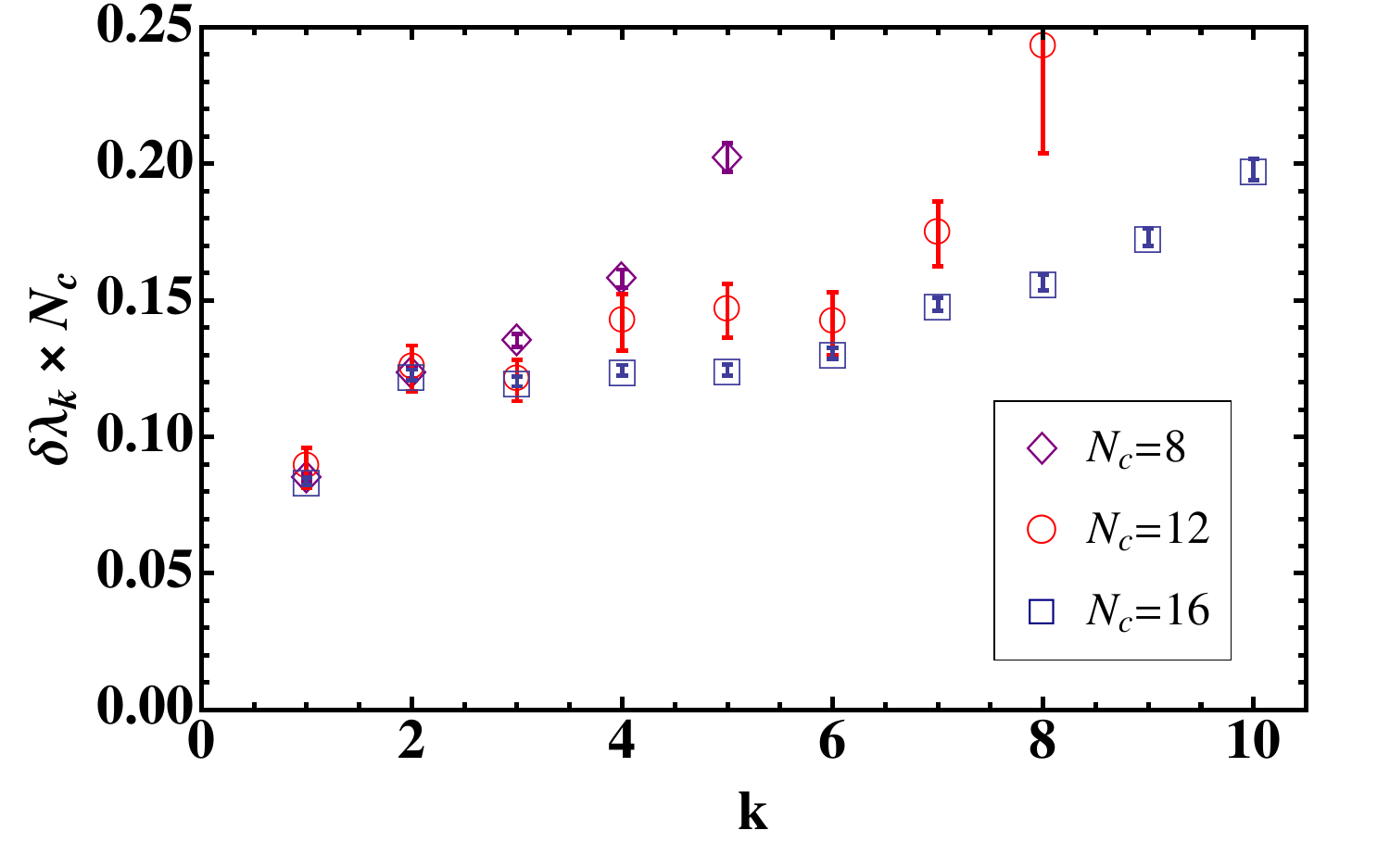}
\caption{%
(Left) Low-lying Dirac spectrum for $N_c=16$, $b=0.5$ and $\kappa=0.09$, where $z$ is the rescaled Dirac eigenvalue obtained by matching the ensemble average of the lowest eigenvalue with the expectation value of the lowest eigenvalue in $\chi$RMT for the ChGSE. Colored bars represent the histograms of individual eigenvalues, while the empty bars represent the accumulated histogram. 
(Right) Dirac eigenvalue spacings mutiplied by $N_c$ for $N_c=8, 12, 16$, where $\delta \lambda_k=\lambda_{k}-\lambda_{k-1}$ and $\lambda_0=0$. 
}
\label{diracspectrum}%
\end{center}
\end{figure}

Our strategy of detecting the S$\chi$SB is to compare microcopic properties of the EK model with those of $\chi$RMT, where the $\chi$RMT limit without losing generality might be achieved by simply taking $m_{probe}=0$ and $N_c\rightarrow\infty$. For this purpose, we calculate the low-lying spectrum of the hermitian overlap-Dirac operator $\mathcal{D}$ for a massless fermion in the adjoint representation.\footnote{The Dirac eigenvalues for adjoint fermions are appearing as conjugate pairs with $2$-fold degeneracy and we take the positive distinct eigenvalues for our numerical results.} 
The operator $\mathcal{D}$ is defined by
\beq
\mathcal{D}=\frac{1}{2}\gamma_5\left[1+\gamma_5~\textrm{sgn}[H_w(-m_0)]\right],
\eeq
where $\gamma_5 H_w(-m_0)$ is the Wilson-Dirac operator and the Wilson mass is set to be negative $-m_0<0$.
Throughout this work, $\chi$RMT predictions are restricted to quenched ($N_f=0$) theory with zero-topological charge. The adjoint QCD belongs to the universal class of the Chiral Gaussian Sympletic Ensemble (ChGSE). However, we also consider two other universal classes, Chiral Gaussian Orthogonal Ensemble (ChGOE) and Chiral Gaussian Unitary Ensemble (ChGUE), in order to make the comparison manifest. As shown in the left figure of \reffig{diracspectrum}, the distribution of low-lying Dirac eigenvalues is in good agreement with that of ChGSE, impling that the chiral symmetry is spontaneously broken.

It might be interesting to find how $m_{probe}$ scales with $N_c$ to extend our study to non-zero mass of the probe fermion or to determine the size of the effective volume in the EK model. 
In the right figure of \reffig{diracspectrum}, we plot the spacings of low-lying Dirac eigenvalues $\lambda_k$ (ensemble averaged) multiplied by $N_c$ for $N_c=8, 12, 16$. 
From the figure, we see a nice agreement between the spacings at up to $k=2$ for $N_c=8$ and at up to $k=3$ for $N_c=12$. 
This agreement implies that the Dirac eigenvalues near zero, which are expected to agree with $\chi$RMT preeiction, scale as $N_c$ and thus $\alpha=1$. 
We also found that the spectral density $\rho(z)$ is nearly zero at between the $(N_c-1)$th and $N_c$th eigenvalues. 

A possible explanation of our founding for the $N_c$-scaling of Dirac eigenvalues may rely on the perturbative analysis in the background of diagonal Wilson lines \cite{Azeyanagi}. 
In a compact space and at weak coupling, one cannot gauge away zero-momentum modes and thus the low-lying Dirac spectrum might be determined by the zero modes. For fermion in the adjoint representation, the number of zero modes of the Wilson lines is $(N_c-1)$ while the total degrees of freedom is $N_c^2$. As a result, the low-lying Dirac eigenvalues should scale as $N_c$, which is consistent with our numerical results. 
Interestingly, our counting of zero modes also agrees with the position at which $\rho(z)\sim0$. 

\section{Conclusion and outlook}

In this proceeding, we discussed about the 't Hooft limit and $\chi$RMT limit in large-$N_c$ gauge theories: in general two limits are not compatible and the large-$N_c$ equivalences do not hold in $\chi$RMT limit. In spite of the difference of these two limits, the S$\chi$SB can be detected by taking an indirect path from $\chi$RMT limit of EK model to the large-volume lattice theory in the 't Hooft limit. As a numerical demonstration, we performed lattice simulations of $SU(N_c)$ gauge theory with two heavy adjoint fermions on a $2^4$ lattice. After confirming that the volume equivalence is valid (unbroken center symmetry), we found that the distributions of low-lying Dirac eigenvalues are in good agreement with the $\chi$RMT prediction and thus the chiral symmetry of the quenched QCD is spontaneously broken.  In the near future, we hope to determine whether the large-$N_c$ dynamical two-flavor adjoint QCD goes through the S$\chi$SB or not, with an application to walking Technicolor theories in mind.

\section{Acknowledgements}

The authors would like to thank S. Hashimoto, A. Hietanen, M. Kieburg, M. Koren, S. Sharpe, M. Tezuka, J. Verbaarschot and N. Yamamoto for stimulating discussions and comments. The numerical calculations were carried out on cluster at KEK. 
This work is supported in part by the Grant-in-Aid for Scientific Research of the Japanese Ministry of Education, Culture, Sports, Science and Technology and JSPS (Nos. 20105002,20105005,and 22740183).

\bibliographystyle{ieeetr}

\end{document}